# Surface Acoustic Wave Amplification by DC-Voltage Supplied to Graphene Film

Z. Insepov[1,2,a)], E. Emelin[3], O. Kononenko[3], D.V. Roshchupkin[3], K.B. Tnyshtykbayev[1], and K.A. Baigarin[1]

[1] *Nazarbayev University Research and Innovation System, 53 Kabanbay Batyr St, Astana, 010000, Kazakhstan*

[2] *Purdue University, 500 Central Drive, West Lafayette, 47907, IN USA*

[3] *Institute of Microelectronics Technology and High Purity Materials, 6, Academician Ossipyan str, Chernogolovka, Moscow region, 142432, Russia*

Using a high-resolution X-Ray diffraction measurement method, the surface acoustic waves (SAW) propagation in a graphene film on the surface of a $Ca_3TaGa_3Si_2O_{14}$ (CTGS) piezoelectric crystal was investigated, where an external current was driven across the graphene film. Here we show for the first time that the application of the DC field leads to a significant enhancement of the SAW magnitude and, as a result, to amplification of the diffraction satellites. Amplification of 33.2 dB/cm for the satellite +1, and of 13.8 dB/cm for the satellite +2, at 471 MHz has been observed where the external DC voltage of +10V was applied. Amplification of SAW occurs above a DC field much smaller than that of a system using bulk semiconductor. Theoretical estimates are in reasonable agreement with our measurements and analysis of experimental data for other materials.

**I. INTRODUCTION**

The development of acousto-electronics based on surface acoustic waves is determined by the use of new piezoelectric materials and by the possibility of controlling propagation of acoustic waves in crystals. The types of control can either be passive or active. The passive control can be realized by fabrication of thin coatings or metallic gratings on crystal surfaces that reduces the speed of SAW or by using domain structures in ferroelectric and ferroelastic crystals [1-3]. An active control can also be implemented by applying an electric field to a piezoelectric crystal. In ref 4, an acousto-electric current was generated in a graphene film fabricated on a lithium-niobate surface where the induced current was controlled by the input rf power.

Here we show the feasibility of amplification of SAWs by applying an external electric DC voltage to a graphene film deposited on a surface of a lanthanum-gallium silicate family $Ca_3TaGa_3Si_2O_{14}$ (CTGS) group.

**II. SAW DEVICE**

Most of the progress in the development of acousto-electronic applications was based on traditional piezo-electric materials such as lead zirconate titanate PZT [5], LiNiO3 [6] coated with a semiconductive film, or using a piezoelectric semiconductor such as CdS [7]. Graphene has never been used for amplification of SAW in piezoelectric materials before. An opposite process, attenuation of SAW, was fist observed in ref 4.

---

[a)] Author to whom correspondence should be addressed. Electronic mail: zinsepov@purdue.edu

A five-component crystal of lanthanum-gallium silicate family $Ca_3TaGa_3Si_2O_{14}$ (CTGS) crystal grown along the {110} crystallographic axis by the Czochralski method was used as piezoelectric material grown.

Recently, major progress in the development of acousto-electronics has been made using the prospective CTGS crystals that belong to the lanthanum gallium silicate (LGS or langasite) group. These crystals have a space symmetry group 32, similar to that of piezo-quartz $SiO_2$. Unlike the $SiO_2$ crystals, the crystals of the lanthanum gallium silicate group do not experience phase transitions until the melting temperature of ~1500 ºC and have significantly higher values for the coefficients of electro-mechanical coupling. Low speed of SAWs in such crystals allows fabrication of miniature acoustic-electric devices.

Substrates for the experiment were fabricated by making a Y-cut (parallel to the (110) plane) of the crystal, having the size of $0.5 \times 8 \times 16$ mm$^3$. Two interdigital transducers (IDTs) consisting of 50 pairs of electrodes with a 3 μm period were fabricated on the substrate surface for generation of SAW, with a wave length of λ = 6 μm, using an electron beam lithography.

A SAW delay line was used as a subject of the present investigation that was discussed in a previous paper (8). This delay line demonstrated 8 dB insertion loss at 471 MHz. The SAW propagation on a piezo-electric substrate CTGS leads to a sinusoidal modulation of the crystal lattice and to formation of an atomic density difference between the minimums and maximums of the acoustic waves due to the lattice period modulation. Such modulations of the lattice parameter can be registered by the X-ray diffraction method used in this work.

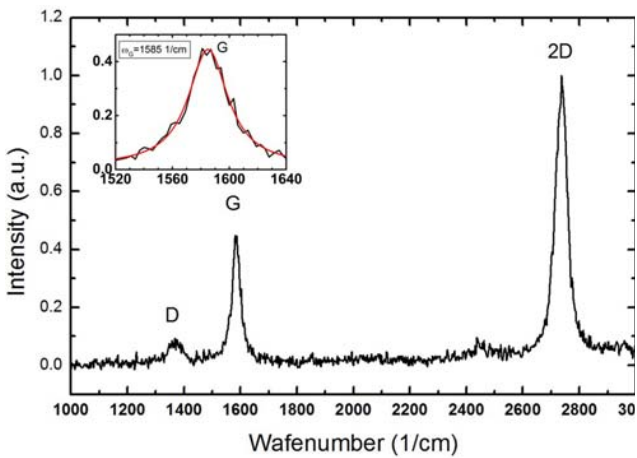

Figure 1. Raman spectra of graphene formed on a CTGS substrate showing the characteristic D, G, and 2D peaks. The ratio I2D/IG = 2.23 indicates that the graphene film consists of few-layers of graphene (FLG).

A graphene film was synthesized on a Ni(111) surface by CVD synthesis from acetylene ($C_2H_2$) and then was transferred to the area between two IDTs on CTGS crystal surface [9].

Figure 1 shows the Raman spectra of the graphene film fabricated on a CTGS crystal surface, with characteristic D-, G-, and 2D-peaks, where irradiation was carried using a blue laser with the wavelength of 488 nm. The intensity ratio for two peaks I2D/IG = 2.23 shows that the fabricated carbon film consists of a few-layers of graphene (FLG) (10).

According to a detailed study of graphene phonon dispersion spectra [11], the location and the shape of the G-peak are important characteristics that determine the layer structure of the deposited carbon film, including the number of layers of graphene.

The inset in Figure 1 shows the Raman spectra of the G peak fitted by a single lorentzian curve, with the maximum at the wavenumber of $\omega_G$ = 1585 1/cm. This maximum value can be obtained from the following equation [11]:

$$\omega_G = 1581.6 + 11/(1+n^{1.6}), 1/cm, \quad (1)$$

Where $\omega_G$ is the wavenumber of the G-peak's maximum, $n$ is the average number of graphene monolayers. According to equation (1), this wavenumber gives $n$ = 2.2. Hence, the carbon film consists of two monolayers of graphene.

Three Al electrodes were fabricated on the top graphene surface by electron beam lithography, with a width of 10 μm and the distances of 300 μm, for application of the DC voltage according to the schematically shown in Fig. 2

### III. EXPERIMENTAL SET-UP

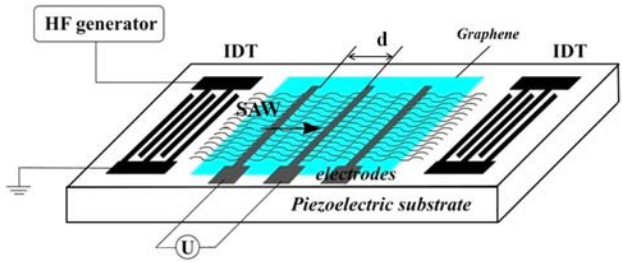

Figure 2. SAW delay time line, with a graphene film fabricated on the surface of piezoelectric substrate. Three Al electrodes were fabricated on the top of graphene by lithography for applying a DC voltage U, d = 300 μm.

Present studies on the possibility of controlling surface acoustic waves with the DC voltage applied to the graphene film, were conducted by using a three-crystal X-Ray diffractometer schematically shown in Figure 3.

The measurements were carried out using a 4-Bounce Bruker D8 Discover X-Ray diffractometer. The X-rays were collimated by an input slit with the width of 100 μm. An X-Ray tube with a rotating copper anode (radiation CuK$_{\alpha 1}$, the wavelength was λ = 1.54 Å). As a double-crystal monochromator, Ge (220) crystals were used with dual reflection. After passing the monochromator, the X-Ray radiation encounters at the Bragg's angle with a Y-cut of the CTGS crystal, modulated by SAW, with a wavelength of λ = 6 μm.

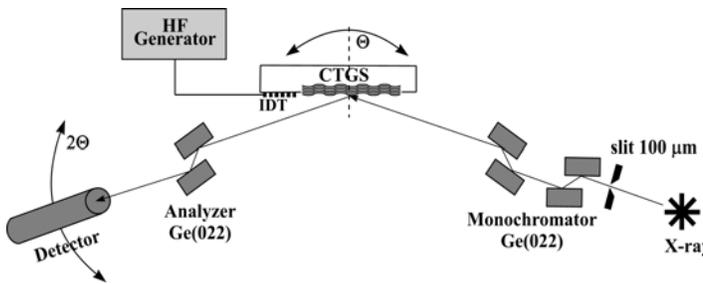

Figure 3. Experimental set-up of a three-crystal X-ray diffractometer.

After the acoustically modulated crystal surface, the X-Rays fall to the analyzing crystal and then registered with a standard NaI scintillation detector. The presence of the sinusoidal modulation of crystal lattice triggers generation of diffraction satellites on the rocking curves, on both sides of the central Bragg's peak. The angular divergence between the diffraction satellites on the rocking curve is determined as:

$$\delta\Theta = \delta/\lambda, \tag{2}$$

Where $\delta$ is the distance between the atomic planes, and $\lambda$ – the wavelength of SAW.

## IV. EXPERIMENTAL RESULTS

The rocking curves of an acoustically modulated CTGS crystal were measured in order to study the effect of application of the DC voltage on the graphene film placed on the top of a piezoelectric crystal during SAW propagation.

Figure 4 represents the rocking curves of a CTGS crystal measured with and without generation of SAWs on the crystal surface. The resonant frequency of SAW with the wavelength of $\lambda = 6$ μm is $f = 471$ MHz, which corresponds to the transversal surface acoustic wave propagation velocity along the X-axis of $v = \lambda \cdot f = 2826$ m/s. To generate the SAW, an rf AC voltage with the amplitude of 15 V was applied to the input IDT.

If the input signal is applied to the IDT, four diffraction satellites will be located on each side around the central Bragg's peak (m = 0). The angular divergence between the diffraction satellites is $\delta\Theta = 0.003°$, which agrees with the value calculated with eq. (2).

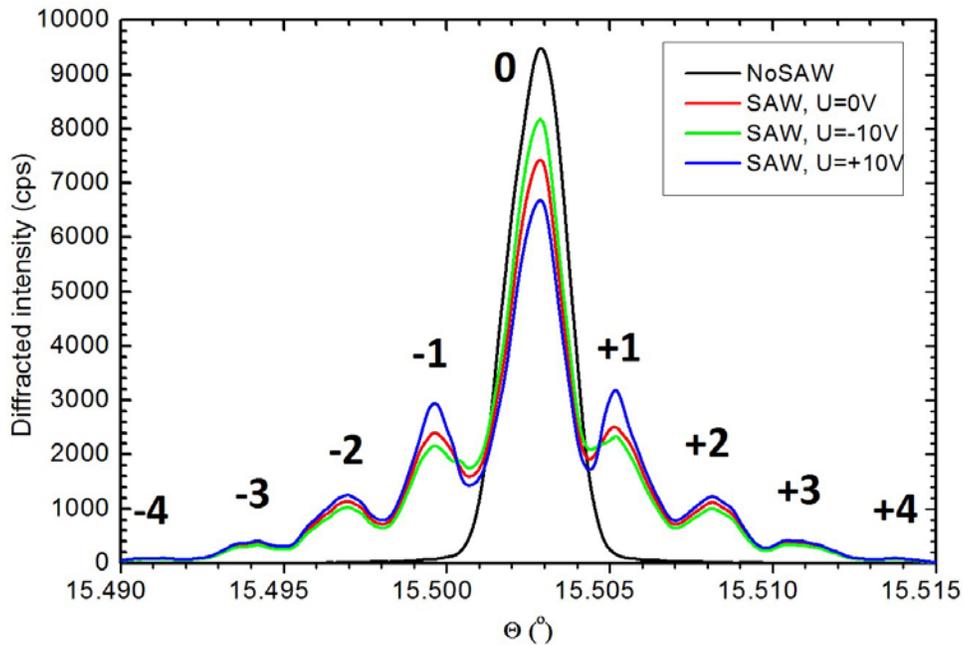

Figure 4. Rocking curves of acoustically modulated Y-cut of a CTGS crystal, $\lambda = 6$ μm: the black curve corresponds to the rocking curve without SAW propagation; the red curve – SAW propagation at zero DC voltage (U = 0 V); the blue line – SAW propagation at a DC voltage of U = +10V; the green line – SAW propagation at a voltage of U = -10V. (The electric field E = U/d = ±333 V/cm)

Figure 4 shows that the value of a DC voltage applied to two neighboring electrodes (see e.g. Figure 2) does not make noticeable changes to the speed of SAWs, as well as to the SAW wave length, since the angular divergence between the diffraction satellites is constant.

However, the DC voltage makes remarkable changes to the intensities of the diffraction peaks, modifying the magnitudes of SAW satellites. Figure 4 shows the rocking curve results of acoustically modulated Y-cut of a CTGS crystal. As it is seen in this figure, application of a DC voltage of U = +10V increases the intensity of the satellites (m = ±1, ±2) and reduces the magnitude of the central peak (m = 0) of diffraction, which is caused by the increase of the magnitude of SAW.

Therefore, amplification of SAW caused by a DC voltage applied to graphene that produces electric current in the direction of the sound wave was observed for the first time in this experiment.

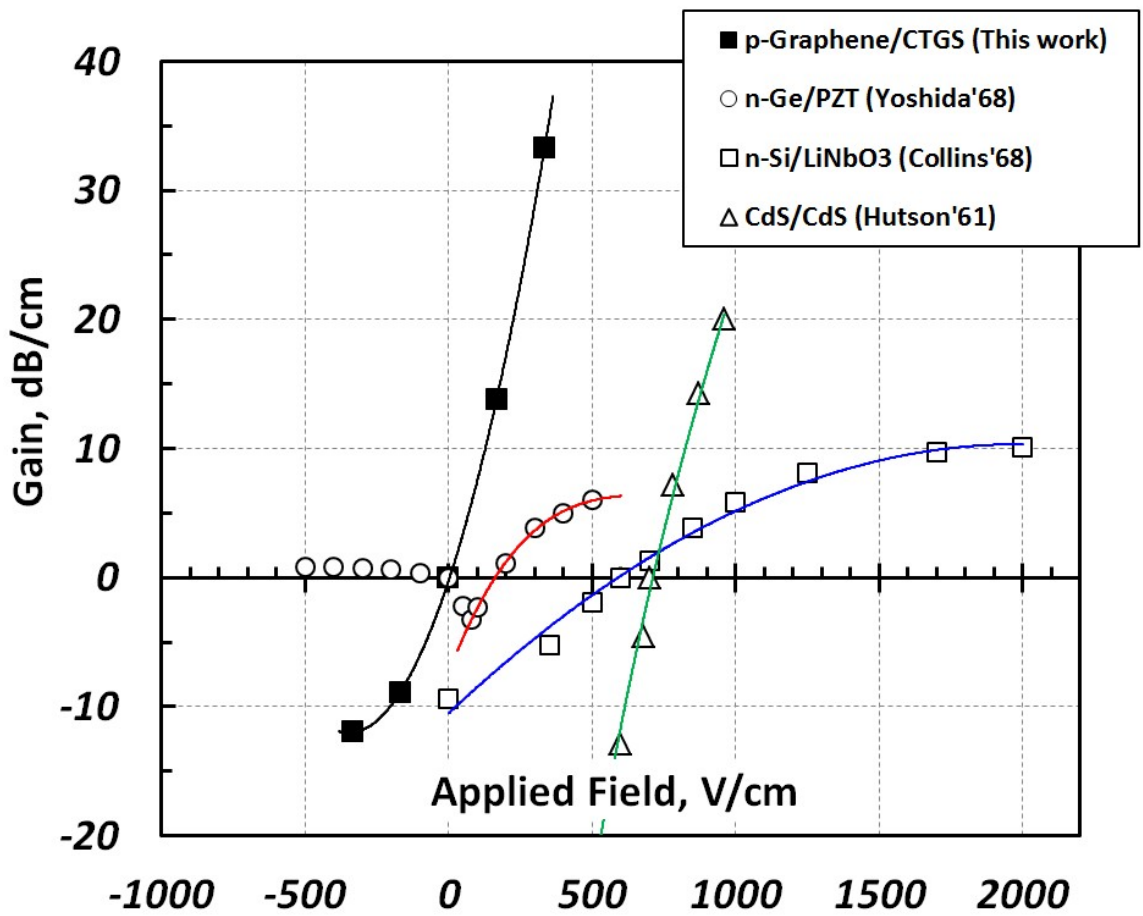

Figure 5. Full symbols are the results for the SAW net gain for 1st satellite of X-Ray diffraction per cm measured in this work for a CTGS piezoelectric coated with graphene: ■ - our results. Open symbols are the experimental data from the following works: ○ - n-Ge/P.Z.T (5) □ - n-Si/LiNbO3 (6), △ - n-CdS/CdS (7).

Application of negative voltages to the graphene film lead to an increase of the intensity of the zeroes Bragg's peak and hence, to a decrease of the intensity of the diffraction satellites, which is caused by the decrease of the SAW magnitude. Figure 4 shows that the SAW signal attenuation at U = -10V. Therefore, application of a drift DC voltage to the graphene film placed on the top of a piezoelectric CTGS crystal allows amplification of the magnitudes of SAW. The amplification effect on graphene was observed in our work for the first time.

Figure 5 shows our experimental results (solid symbols) obtained at various DC voltages on graphene in an interval between -10V and +10V. The experimental data points were measured as amplification of the 1st diffraction satellite magnitude of SAW at different voltages applied to the graphene film. Since the graphene was grown in the atmosphere, the film had a p-type which was confirmed by the Hall measurements. The SAW signal attenuates in the region U < 0, and the acoustic signal amplifies in the region with U > 0. Therefore, Figure 5 shows only one type of carriers, in our case holes, that was available in graphene for interaction with the electro-magnetic field generated by SAW on the piezoelectric CTGS surface.

The effect of SAW amplification was not studied in the systems built by graphene before. In non-graphene systems, the SAW amplification effect was previously studied experimentally starting from the beginning of 1960s [5-7]; and by theory in refs 12-14. Comparison of the experimental data for two non-graphene acousto-electric systems fabricated by a piezo-electric in an acoustic contact with a semiconductor is shown in Figure 5. Open circles show the data for SAW amplification measured with a polarized PZT ceramic slice with comb electrodes on both sides, with a tuning frequency of 20 MHz, and a polished n-Ge wafer (0.5 cm of length, 50 Ω·cm) in which the carriers drift [5]. Open squares correspond to a high-performance delay line using Y-cut LiNbO3 with in an acoustic contact with an n-Si bar (200 Ω·cm) mounted on the top in ref 6. The length of the acoustic contact in Si was 1.52 cm. Ultrasonic amplification was studied by applying a pulsed rf signal, with a frequency of 108 MHz, synchronized with an illumination of n-Si bar with a laser radiation that generated photoelectrons. The DC current was flown in the Si bar simultaneously with the rf input. The contribution of holes in [6] was not taken into account. A piezo-electric semiconductor CdS crystal (0.7cm of length, shear wave velocity of $2\times10^5$ cm/s, electron mobility of 285 cm2/Vs) was used in [7] at two rf frequencies, 15 and 45 MHz, where the DC current was flown via the same crystal during its illumination by a mercury radiation line.

Since our experimental technique is based on X-ray diffraction measurements, it cannot register the losses of insertion of IDTs, contacts and the graphene film. Therefore, the experimental amplification data from literature [5-7] were corrected by deducting the insertion losses that were indicated in the above papers, thus making them comparable to our measurements.

Since the experimental amplification data greatly vary depending on the length of the active acoustic contact of conducting area with that of piezo-electrics, all experimental results were normalized per unit length.

## V. DISCUSSION AND CONCLUSION

The experimental data of the 1st X-ray diffraction satellite magnitude change was generated by applying the DC voltage on graphene film and are collected in Figure 5. Attenuation and amplification of SAW in piezo-electric materials acoustically contacting with a semiconducting film or in a semiconducting CdS piezoelectric substrate were studied experimentally, where SAW propagation and electric flow were physically occurring in PZT [5], LiNbO$_3$ [6] or a semiconducting piezoelectric CdS [7]. The onset of amplification for the gain was reported at the electric fields above 150 V/cm for PZT [5], 600 V/cm for LiNbO$_3$ [6], and of 700 V/cm for CdS [7]. One of the reasons for such high electric fields could be a narrow air gap between the piezo-crystal and the semiconductor that could strongly reduce the electric field generated by the piezo-electric. The narrow air gap was specifically introduced for elimination of the surface stresses in the piezo-material due to coating [5]. Another reason of the higher electric field threshold in the traditional systems could be a relative low charge carrier's mobility in all three semiconductors Ge, Si and CdS, compared to that of graphene. The conductivity type of the CVD few-layer graphene was obtained for our samples by Hall measurements that showed the p-type with the hole mobility of $\mu_h$ = 4800 cm$^2$/ V·s.

Our preliminary X-Ray diffraction data for positive (amplifying) and negative (attenuating) fields shown in figure 5 were smoothly connected by a spline line that crosses the X-axis very close to zero, and therefore, does not show any threshold for the electric DC field. Therefore, they are much smaller than 150 V/cm or ~ 600-700 V/cm in the conventional systems based on coatings by semiconductor films [5-7].

Our experimental results for SAW amplification can be explained by interaction of the electric current drifting in graphene under the electric field by applied DC voltage, with the longitudinal electro-acoustic fields generated in the CTGS piezoelectric by SAW propagation [12-14].

This threshold is explained by a resonant character of the sound amplification effect. The amplification effect exists only if: i) the velocity of the drifting charge carriers is in the same directions as the SAW propagation, and ii) it is equal or higher than the phase velocity of the shear acoustic wave vph. The acoustic signal attenuates if the carriers are drifting in an opposite direction. Therefore, if the drifting electric charges reach the supersonic speed in the film, they start to generate acoustic phonons, in an analogy with the Landau damping in plasma [15]. According to this model (see, e.g. [12-14]), the electro-magnetic wave traveling with a phase velocity $v_{ph}$ strongly interacts with the charge carriers having velocities collinear and close to $v_{ph}$ (the resonant particles). As a result, the particles having velocities higher than $v_{ph}$ will be transferring their kinetic energy to the wave, thus amplifying the wave, while those which are slower will be accelerated by wave and thus, attenuate the wave. The drift velocity of electrons under a DC electric field E can be obtained as follows:

$v_E = \mu_h E,$ (3)

Where $v_E$ is the carrier drift velocity, and the electric field E=U/d, the hole mobility $\mu_h$ = 4800 cm$^2$/V·s of the graphene film was obtained in our Hall measurements, U is the DC voltage, and d = 300 μm, - the distance between the Al electrodes on the graphene film. The threshold condition $v_E = v_s$, where $v_s$ = 2826 m/s is the sound speed in the CTGS piezoelectric that determines the threshold field. Using the above parameters, the threshold voltage was obtained as follows: $E_{th} = v_s/\mu$ = 58.9 V/cm.

TABLE 1

| Reference | Material | Sound speed, cm/s | Electron mobility, cm2/V/s | Hole mobility, cm2/V/s | Threshold field via eq. (3), V/cm | Experimental threshold from this work and refs. |
|---|---|---|---|---|---|---|
| This work | graphene | 2.83E+05 | N/A | 4.80E+03 | 58.9 | 0 |
| (16) | Ge | 5.40E+05 | 3900 | 1900 | 138. 5 | 150 (5) |
| (16) | Si | 8.43E+05 | 1500 | 471 | 562.2 | 600 (6) |
| (7) | CdS | 2.00E+05 | 285 | 40 | 701.8 | 600 (7) |

The experimental data for graphene and other semiconductors are compared in Table 1 with the theoretical data calculated using eq. 3. This comparison shows good agreement between the experimental and theoretical data calculated with eq. 3 for such a wide variety of different elastic and electronic properties.

Therefore, our present results show that a higher mobility of graphene films gives an opportunity of making micrometer-sized acousto-electric amplifiers that operates at much lower DC voltages. This enables stronger amplification of ultrasound waves at small sizes, for future MEMS devices, that are impossible to reach with conventional semiconductors, and that are important for biomedical and nanofluidics applications.

A significant engineering problem arises in the development of a novel solid-state air or liquid pumping technology that brings high flux rates in nano and micro channels. Actuation of a fluid flow in micro-capillaries has fundamental interest in the areas of nanofluidics [17]. The demonstration of the amplification effect using graphene films facilitates industrial applications of SAW for actuation of liquid and gas flows at nano and micro scales.

In conclusion, a feasibility of controlling the SAW propagation process caused by applying a drift voltage applied to a graphene film placed on the top of a CTGS piezo-electric crystal was demonstrated in this article. Application of the drift voltage allows changing the magnitude of SAW. It should be pointed out that the graphene film is very light and, therefore, does not influence the SAW propagation process and introduces much less signal loss than that of conventional coatings.

**ACKNOWLEDGMENT**

This work was supported in part by the Nazarbayev University World Science Stars program, under grant No. 031-2013 of 12/3/2013.